\documentclass[reprint,amsmath,amssymb,aps,prb]{revtex4-2}

\usepackage{graphicx, float}
\graphicspath{{./figures/}}
\DeclareGraphicsExtensions{.eps,.pdf,.mps,.jpeg,.png}
\usepackage{dcolumn}
\usepackage{siunitx}
\usepackage{bm}
\usepackage[hidelinks,colorlinks=true,allcolors=blue]{hyperref}

\begin{document}

\preprint{APS/123-QED}
\title{Enhancement of supercurrent through ferromagnetic materials by interface engineering}
\author{Swapna Sindhu Mishra}
\author{Robert M. Klaes}
\author{Joshua Willard}
\author{Reza Loloee}
\author{Norman O. Birge}%
\email{birge@msu.edu}
\affiliation{Department of Physics and Astronomy, Michigan State University, East Lansing, MI}
\date{\today}

\begin{abstract}
Josephson junctions containing ferromagnetic materials exhibit interesting physics and show promise as circuit elements for superconducting logic and memory. For memory applications, the properties of the junction should be controllable by changing the magnetic configuration inside the junction. To achieve good magnetic switching properties, one should choose a soft magnetic material such as NiFe (permalloy); however, NiFe exhibits poor supercurrent transmission in Josephson junctions. In this work we put thin layers of Ni on either side of the NiFe and characterize the magnetic behavior and supercurrent transmission properties of the Ni/NiFe/Ni trilayers as a function of Ni and NiFe thicknesses. Using a Ni thickness of 0.4 nm, we find that the magnetic switching behavior of the trilayers is not severely degraded relative to plain NiFe, while the maximum supercurrent in the $\pi$-state of the trilayer Josephson junctions is increased by a factor of four relative to that of NiFe junctions. We speculate that the supercurrent enhancement is due to the different spin-dependent transport properties of the Cu/Ni and Cu/NiFe interfaces.   
\end{abstract}

\maketitle

\section{Introduction}
Josephson junctions containing ferromagnetic materials are the subject of intense study both because of their rich physics \cite{Buzdin2005} and because they show promise for applications in superconducting digital logic and memory \cite{Ryazanov2012,Soloviev2017} and for various superconducting qubit designs for quantum computing \cite{Ioffe1999, Blatter2001, Yamashita2005, Feofanov2010}. Due to the exchange splitting between the majority and minority spin bands in ferromagnetic (F) materials, spin-singlet Cooper pairs undergo rapid phase oscillations and decay in the F layer \cite{Demler1997,Buzdin2005}. As a result of these oscillations, the ground-state phase difference across the junction can be either 0 or $\pi$ depending on the thickness of the F layer \cite{Ryazanov2001,Kontos2002}. Such $\pi$-junctions have been proposed as circuit elements in single-flux-quantum (SFQ) logic and memory circuits of several different kinds \cite{Ustinov2003, Ortlepp2006, Khabipov2010, Kamiya2018,Takeshita2021, Katam2018}. Interest in SFQ circuits has risen in recent years due to their potential for energy-efficient computing \cite{Holmes2013} or for use as an interface to a cryogenic quantum computer \cite{Howington2019}.

For memory applications it would be advantageous to be able to change the properties of a junction by changing the magnetic configuration inside the junction \cite{Krivoruchko2001, Golubov2002}. That can be achieved by inserting two independent F layers inside the junction in a ``pseudo spin-valve” configuration, in which the magnetization of one layer remains fixed while the other is free to rotate in a small external field. Using such a scheme, several groups have demonstrated modulation of the critical current amplitude \cite{Bell2004,Baek2014,Qader2014} or the phase state of the junction \cite{Gingrich2016,Dayton2018,Madden2018}. 

The applications mentioned above can be divided into two classes: those where the ferromagnetic Josephson junction acts as a passive phase shifter, always remaining in the supercurrent-carrying state, and those where the critical current is occasionally exceeded, causing the junction to switch momentarily into the voltage state. In the former case, the junction is typically surrounded by conventional superconductor/insulator/superconductor (S/I/S) junctions, which undergo switching during logic or memory read operations. The ferromagnetic (S/F/S) junction must then have larger critical current ($I_c$) than the nearby S/I/S junctions to avoid switching of the S/F/S junction into the voltage state. While one can increase $I_c$ simply by increasing the lateral area of the junction, that is undesirable because it causes the magnetic layers to be in a multi-domain state, which is detrimental to their magnetic switching properties. Instead, it is preferable for the S/F/S junction to have a critical current density, $J_c$, that is much larger than that of the nearby S/I/S junctions. In the latter case where the junction switches into the voltage state during logic operations, one desires instead to have a large $I_cR_N$ product, where $R_N$ is the normal-state resistance of the junction, because the speed at which the junction switches from the supercurrent state to the voltage state is proportional to $I_cR_N$. That can be achieved by inserting a thin insulating barrier inside the junction in tandem with an extra superconducting layer to make a so-called S/I/s/F/S junction \cite{Larkin2012,Vernik2013}. In this work we are interested in the first class of junctions; hence our goal is to optimize $J_c$ as opposed to $I_cR_N$.

The Josephson junctions used in many of the junction-modulation demonstrations mentioned earlier \cite{Bell2004,Baek2014,Qader2014, Gingrich2016, Dayton2018} used permalloy (a Ni-Fe alloy with approximately 80\% Ni) as the free layer, because it is a soft magnetic material with low coercivity. Unfortunately, supercurrent transmission through permalloy is low compared to, for example, pure Ni \cite{Robinson2006,Glick2017,Baek2018}. But the latter has high coercivity, not suitable for a free layer, and was in fact used as the fixed magnetic layer in most of the demonstrations mentioned above \cite{Baek2014,Qader2014, Gingrich2016, Dayton2018}. The relative size of the supercurrent transmission through permalloy (henceforth called ``NiFe”) and Ni has not been addressed theoretically, because nearly all the theory papers describing S/F/S Josephson junctions treat the band structure of the F materials in an oversimplified way \cite{Buzdin2005}. (We will discuss the one exception \cite{Ness2022} later.) To achieve our goal of increasing the critical current density, $J_c$, we must turn to empirical knowledge gained either from previous studies of S/F/S junctions (such as those cited above), or from related studies of spin-polarized transport in the normal state. For the latter, we are very fortunate. Through studies of Giant Magnetoresistance (GMR) of metallic multilayers, the spin-dependent transport properties of many ferromagnetic materials and their interfaces with normal metals have been measured by Bass and Pratt and their collaborators at Michigan State University, and tabulated in a review article \cite{Bass2016}. It is known, for example, that the minority-spin electrons in NiFe have a very short mean free path and also have poor transmission at a NiFe/Cu interface \cite{Vila2000}. Those two properties may largely be responsible for the small supercurrent in NiFe-containing Josephson junctions, as well as its rapid decay with increasing NiFe thickness. The GMR work cited above has shown that both the boundary resistance and spin-scattering asymmetry are smaller for the Cu/Ni interface than for the Cu/NiFe interface \cite{Bass2016}. Based on those results, our strategy in this work is to ameliorate the poor interface transmission by inserting thin Ni layers at each Cu/NiFe interface -- i.e. by replacing NiFe with a Ni/NiFe/Ni trilayer. The results of this strategy are promising: we find that we can increase the supercurrent transmission through NiFe by about a factor of 4 by using 0.4 nm of Ni on each side of the trilayer. The magnetic switching behavior of the trilayer is only slightly degraded compared to that of a NiFe film of comparable thickness.

\section{Fabrication and Measurement}
\subsection{Thin films}
Thin multilayer films with structure base/Ni(0.4)/NiFe($d_{\mathrm{NiFe}}$)/Ni(0.4)/cap (layer thicknesses in nanometers) were deposited on 0.5 inch$\times$0.5 inch Si chips by dc triode magnetron sputtering. The base layer used was [Nb(25)/Al(2.4)]$_3$/Nb(20)/Cu(2) to match the base electrode of our Josephson junctions. The deposited films were capped with Cu(2)/Nb(5) to prevent oxidation. Our NiFe sputtering target has a nominal composition of Ni$_{81}$Fe$_{19}$, while energy dispersive X-ray (EDX) analysis of thick sputtered films suggest a film composition closer to Ni$_{82}$Fe$_{18}$. Sputtering deposition was performed at an Ar pressure of 0.3 Pa and a substrate temperature of 250 K. The base pressure of the sputtering chamber was $4 \times 10^{-6}$ Pa. The NiFe layer thickness, $d_{\mathrm{NiFe}}$ was varied from 0.4 to 3.2 nm in steps of 0.4 nm. For comparison, we sputtered thin NiFe($d_{\mathrm{NiFe}}$) samples with the same $d_{\mathrm{NiFe}}$. We also sputtered Ni(0.2)/NiFe($d_{\mathrm{NiFe}}$)/Ni(0.2) trilayers with the same base and capping layers with $d_{\mathrm{NiFe}}$ varying from 0.4 to 1.8 nm in steps of 0.2 nm. All depositions were performed in the presence of a small magnetic field of about 20 mT to align the magnetocrystalline anisotropy of the NiFe in a known direction. The target voltages and currents used for different materials are listed in \cite{Glick2017Thesis}. The sputter rates are highly stable and are measured every few minutes to ensure there is no variation during the process. The stability of the sputter rates along with the computer control of the deposition times make the deposition of extremely thin layers highly reproducible from sample to sample and across different runs. Film thicknesses for several materials have been verified using low-angle X-ray reflection measurements. Polarized neutron reflectometry studies of superlattices also confirm that the thicknesses are very close to the nominal deposition thicknesses \cite{Quarterman2020}. 

The moment vs field measurements for all sets of thin films were performed using a SQUID-based Vibrating Sample Magnetometer (VSM) at a temperature of 10 K.

\subsection{Josephson junctions}
The detailed fabrication process for ferromagnetic Josephson junctions has been published previously \cite{Glick2017}. The bottom lead photo-lithographic stencil was patterned on a Si substrate, and then [Nb(25)/Al(2.4)]$_3$/Nb(20)/Cu(2)/F/Cu(2)/Nb(5)/Au(10) was sputtered where F denotes the set of ferromagnetic layers. We deposited three different sets of junctions: in the first set, the F layers consisted of Ni(0.4)/NiFe($d_{\mathrm{NiFe}}$)/Ni(0.4) where $d_{\mathrm{NiFe}}$ was varied from 0.4 to 3.4 nm in steps of 0.2 nm; in the second set, the F layers were Ni(0.2)/NiFe($d_{\mathrm{NiFe}}$)/Ni(0.2) where $d_{\mathrm{NiFe}}$ was varied from 0.2 to 1.6 nm in steps of 0.2 nm; in the third set the F layers were NiFe($d_{\mathrm{NiFe}}$) where $d_{\mathrm{NiFe}}$ was varied from 1.0 to 3.7 nm in steps of 0.1 nm. The depositions were performed at similar substrate temperature and Ar pressure as our thin films described earlier. Elliptical junctions with lateral dimensions of \SI{1.25}{\micro \meter}$\times$\SI{0.5}{\micro \meter} and major axis oriented along the magnetic easy axis defined by the magnetocrystalline anisotropy were then patterned by e-beam lithography with a negative ma-N2401 e-beam resist followed by ion-milling. The area surrounding the junctions was then covered with $\mathrm{SiO_x}$ \textit{in-situ} to avoid shorting between the bottom and top superconducting electrodes. The e-beam resist was then removed, and the top lead stencil was patterned using photolithography. After 5 nm of the previous Au(10) capping layer was ion milled \textit{in-situ}, the top Nb(150)/Au(10) superconducting electrodes were deposited by sputtering.

Josephson junctions were mounted on a probe with a built-in superconducting magnet and inserted inside a liquid helium dewar for transport measurements at 4.2 K. The samples were first initialized in a magnetic field of 0.2 T applied along the magnetic easy axis (elliptical major axis) to fully saturate the magnetic layers. Then $I$-$V$ curves were measured as a function of field, in fields up to 0.1 T in both directions.

\section{Results}
\subsection{Thin film magnetics}
\begin{figure}[!htbp]
\includegraphics[width=\linewidth]{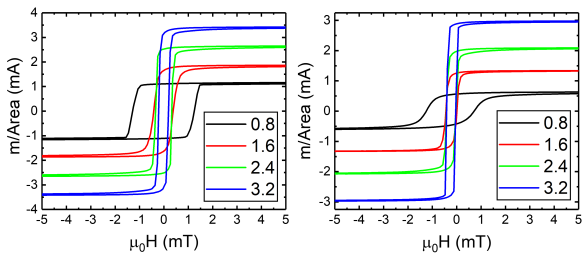}
\caption{Moment/Area ($m/\mathrm{Area}$) vs field ($H$) for selected Ni(0.4)/NiFe($d_{\mathrm{NiFe}}$)/Ni(0.4) (left) and NiFe($d_{\mathrm{NiFe}}$) (right) samples measured at $T = 10$ K.  The values of $d_{\mathrm{NiFe}}$ are shown in both panels.}
\label{fig:MvH}
\centering
\end{figure}

Fig \ref{fig:MvH} shows the moment per unit area vs field for a selected set of Ni(0.4)/NiFe($d_{\mathrm{NiFe}}$)/Ni(0.4) and NiFe($d_{\mathrm{NiFe}}$) samples. As expected, each trilayer behaves as a single ferromagnetic film due to the strong exchange coupling across the interfaces. The coercivities of the samples with Ni are higher than those of plain NiFe samples, but they are still relatively low. The switching profile is also relatively sharp even though we see traces of slow Ni switching near the closing of the hysteresis loops. Note that the total thickness of each sample in the left panel is 0.8 nm greater than the corresponding sample in the right panel due to the Ni layers, so one should compare a Ni/NiFe/Ni sample with a NiFe sample that is $\approx 0.8$ nm thicker.

\begin{figure}[!htbp]
\includegraphics[width=\linewidth]{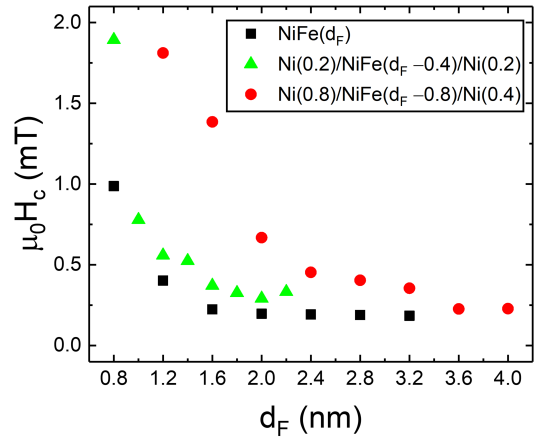}
\caption{Coercivity ($H_c$) vs total F-layer thickness ($d_F$) for NiFe($d_{F}$) (black squares), Ni(0.2)/NiFe($d_{F}-0.4$)/Ni(0.2) (green triangles) and Ni(0.4)/NiFe($d_{F}-0.8$)/Ni(0.4) (red circles) samples measured at $T = 10$ K.}
\label{fig:Hc_Comparison}
\centering
\end{figure}

Fig. \ref{fig:Hc_Comparison} shows the coercivities vs total ferromagnetic (F) layer thickness for NiFe($d_{F}$), Ni(0.2)/NiFe($d_{F}$-0.4)/Ni(0.2) and Ni(0.4)/NiFe($d_{F}-0.8$)/Ni(0.4) samples. We chose the total F-layer thickness instead of the NiFe thickness as the x-axis to better compare these samples with varying Ni thicknesses. (This method makes more sense than comparing samples with the same NiFe thicknesses, but it is not perfect since Ni and NiFe have different magnetizations.) The figure shows that addition of the two thin Ni layers increases the coercivity of NiFe considerably in the thinnest samples, but only by a small amount in the thicker samples. 

Assuming that magnetization reversal in thin films proceeds largely by domain wall motion, the large coercivities of the thinnest films imply that domain wall pinning is dominated by surface effects. For applications in sub-micron Josephson junctions, it is advisable to make the junctions small enough so that the nanomagnets are single-domain and switch via coherent rotation of the magnetization (Stoner-Wohlfarth switching). In that case, shape anisotropy supplies an additional energy barrier to switching. As a result, the coercive fields shown in Fig. \ref{fig:Hc_Comparison} should be viewed as lower bounds to the switching fields of the magnetic layers in the Josephson junctions.

\begin{figure}[!htbp]
\includegraphics[width=\linewidth]{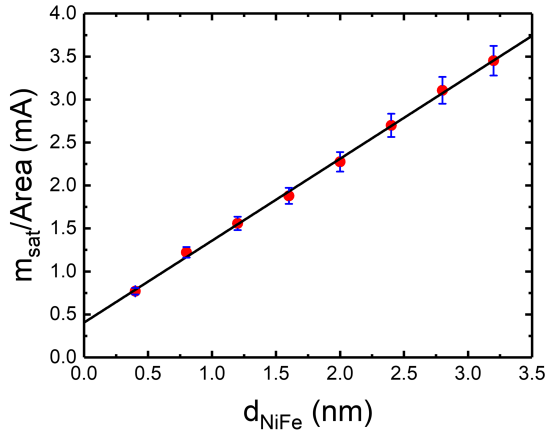}
\caption{Saturation moment per unit area vs NiFe thickness for Ni(0.4)/NiFe($d_{\mathrm{NiFe}}$)/Ni(0.4) samples measured at $T = 10$ K.  The line is linear fit to the data discussed in the text.}
\label{fig:Msat_NiNiFe}
\centering
\end{figure}

Fig. \ref{fig:Msat_NiNiFe} shows the saturation moment/area vs $d_{\mathrm{NiFe}}$ for Ni(0.4)/NiFe($d_{\mathrm{NiFe}}$)/Ni(0.4) samples. The error bars represent 5\% uncertainty attributed to the area measurement method.  (The film deposition does not cover the edge of the chips completely because of the shape of the sample holders; this small non-deposited area is hard to distinguish with 100\% accuracy under the microscope during area measurements.) By fitting these points to a straight line, we determine the value of $M_{\mathrm{NiFe}}$ from the slope to be $952 \pm 14$ kA/m and the value of $M_{\mathrm{Ni}}$ from the intercept to be $507 \pm 36$ kA/m. Our Ni value is very close to the nominal low-temperature magnetization value for bulk Ni of 510 kA/m \cite{OHandley2000}, implying that there are minimal magnetic ``dead layers" at the Cu/Ni and Ni/NiFe interfaces. Our NiFe value is somewhat higher than expected; O'Handley \cite{OHandley2000} lists the low-temperature magnetization of Ni$_{80}$Fe$_{20}$ as 930 kA/m, so the magnetization for our concentration should be somewhat less. We attribute the discrepancy to the uncertainty in the sample areas and the small thickness range explored. 

\subsection{Josephson junction transport}
Josephson junctions containing ferromagnetic materials exhibit overdamped dynamics in the absence of an insulating barrier in the junction. The $I-V$ curves of these junctions can be fit to the Resistively Shunted Junction model: \cite{BaronePaterno1982}
\begin{equation}
    V = \mathrm{sign}(I) R_N \mathrm{Re}\left\{\sqrt{I^2-I_c^2}\right\}
\end{equation}
where the critical current, $I_c$ and the normal-state resistance, $R_N$ can be estimated from fitting the experimental data with the above equation.

\begin{figure}[!htbp]
\includegraphics[width=\linewidth]{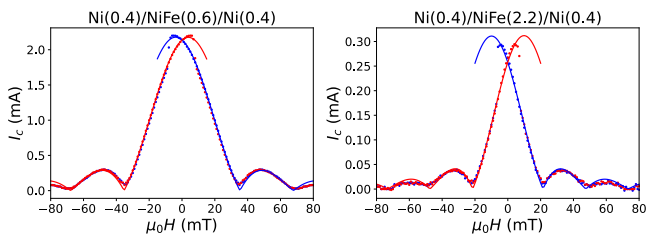}
\caption{Critical current vs magnetic field for Josephson junctions containing Ni(0.4)/NiFe(0.6)/Ni(0.4) (left) and Ni(0.4)/NiFe(2.2)/Ni(0.4) (right). The red and blue points represent data taken during field upsweep and downsweep, respectively. The solid lines are fits to Eqn. \ref{Eqn:Airy}.}
\label{fig:Fraunhofers}
\centering
\end{figure}

Fig. \ref{fig:Fraunhofers} shows the dependence of $I_c$ on varying magnetic field $H$ applied along the long-axis of the elliptical junctions for two representative samples: Ni(0.4)/NiFe(0.6)/Ni(0.4) and Ni(0.4)/NiFe(2.2)/Ni(0.4). The blue and red data points were acquired during the field downsweep and upsweep, respectively. 

For elliptical junctions with the field applied along a principal axis, the data are expected to follow an Airy function \cite{BaronePaterno1982}:
\begin{equation} \label{Eqn:Airy}
    I_c(\Phi) = I_{c0} \left| \frac{2 J_1 \left( \frac{\pi \Phi}{\Phi_0} \right) }{\frac{\pi \Phi}{\Phi_0}} \right|
\end{equation}
where $I_{c0}$ is the maximum value of $I_c$, $J_1$ is the Bessel function of the first kind, $\Phi_0 = 2 \times 10^{-15}$ Tm$^2$ is the flux quantum, and the total magnetic flux through the junction is given by:
\begin{equation} \label{Eqn:Flux}
    \Phi = \mu_0 H w (2\lambda_{\mathrm{eff}} + d_N + d_F) + \mu_0 M w d_F
\end{equation}
where $\lambda_{\mathrm{eff}}$ is the effective London penetration depth, $d_N$ is the thickness of the normal layers, $d_F$ is the thickness of the ferromagnetic layers, $M$ is a weighted average of the Ni and NiFe magnetizations, and $w$ is the width of the junctions transverse to the field direction. (Note that Eqn. \ref{Eqn:Flux} neglects the very small demagnetizing field as well as any flux from the F layer that returns inside the junction -- i.e. between the bottom and top S electrodes.)  The solid lines in Fig. \ref{fig:Fraunhofers} are fits of Eqn. \ref{Eqn:Airy} to the experimental data. The center of the Airy pattern is shifted in either direction because of hysteresis arising from the internal magnetization of the ferromagnetic layers in the junction. There is also a sudden drop in $I_c$ around $\pm 10$ mT due to the switch in the direction of the magnetization. The field shift and switching field depend on the magnetic properties of the ferromagnetic layers and vary with NiFe thickness. Due to this sudden switch, the value of $I_{c0}$ in Eqn. \ref{Eqn:Airy} is typically higher than the value realized experimentally and must be extracted from the fit, as shown by the extended fit lines in the figure. We chose these two representative samples to show that all the above properties, including the maximum $I_c$, are dependent on the thickness of the ferromagnetic layer via the flux $\Phi$. The period of the Airy pattern is dependent on the shape and size of the ellipse and should be the same for all samples. However, there is a difference in the periods of the two Airy patterns that is not fully understood, but is partly due to some variation in the junction widths during fabrication, discussed in Appendix A.

\begin{figure*}[!htbp]
\includegraphics[width=\linewidth]{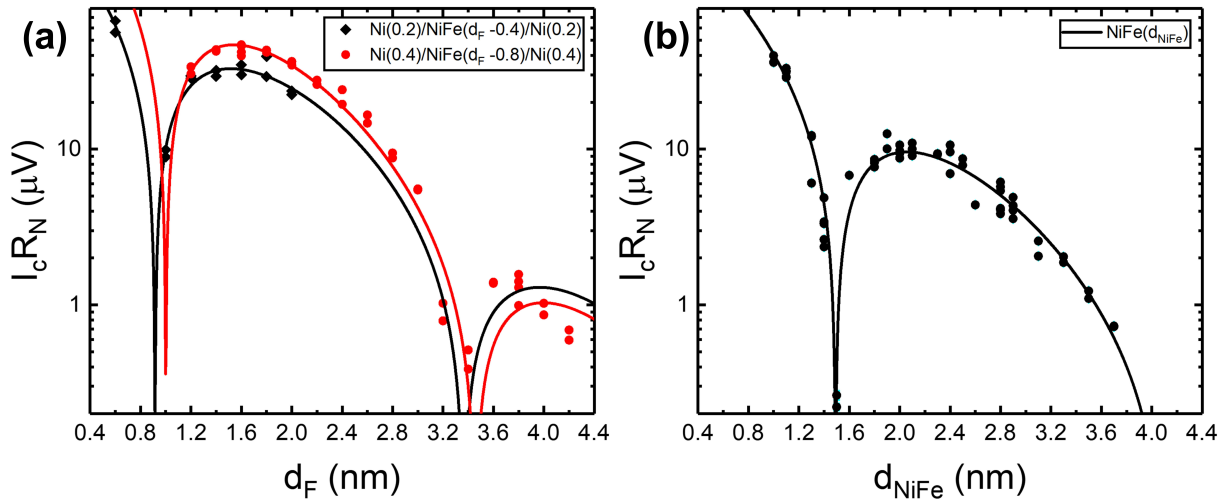}
\caption{(a) $I_c R_N$ vs total F-layer thickness $d_F$ = $d_{\mathrm{NiFe}}+d_{\mathrm{Ni}}$ for Ni(0.4)/NiFe($d_F$-0.8)/Ni(0.4) (red circles), Ni(0.2)/NiFe($d_F$-0.4)/Ni(0.2) (black diamonds) and (b) $I_c R_N$ vs NiFe thickness $d_{\mathrm{NiFe}}$ (right) for NiFe($d_{\mathrm{NiFe}}$). Each data point represents the average of the field upsweep and downsweep data for a single Josephson junction. The solid lines represent the fits to Eqn. \ref{Eqn:Diffusive} discussed in the text. The fits for Ni(0.2)/NiFe($d_F$-0.4)/Ni(0.2) and NiFe($d_{\mathrm{NiFe}}$) sets are extended to the same range as the Ni(0.4)/NiFe/Ni(0.4) data for better comparison. Due to the absence of data points for higher thicknesses, the thicknesses for the second 0-$\pi$ transitions seen in the plot for these two data sets might be unreliable.}
\label{fig:ICRN}
\centering
\end{figure*}

To obtain the maximum critical current density, $J_{c0}$, one would divide the measured value of $I_{c0}$ for each junction by the junction area. Due to the process variation discussed in Appendix A and the difficulty in obtaining accurate values of the individual junction areas, we plot instead the product of $I_c$ times the measured normal state resistance, $R_N$. The $I_c R_N$ product is nearly independent of junction area and serves as a surrogate for $J_c$. Figure \ref{fig:ICRN}(a) shows $I_c R_N$ vs total F-layer thickness $d_F=d_{\mathrm{NiFe}}+d_{\mathrm{Ni}}$ for all the measured Josephson junctions containing either Ni(0.4)/NiFe($d_F$-0.8)/Ni(0.4) or Ni(0.2)/NiFe($d_F$-0.4)/Ni(0.2). For comparison, Fig. \ref{fig:ICRN}(b) shows $I_c R_N$ vs NiFe thickness $d_{\mathrm{NiFe}}$ for a set of junctions containing only a single NiFe layer of thickness $d_{\mathrm{NiFe}}$, without any Ni. Just by looking at the raw data without doing any fitting, one can see immediately that the maximum value of $I_c R_N$ in the $\pi$-state of the junctions shown in panel (a), occurring at $d_F \approx 1.6$ nm, is four to five times larger for the Ni(0.4)/NiFe/Ni(0.4) junctions and about three times larger for the Ni(0.2)/NiFe/Ni(0.2) junctions, compared with the NiFe junctions shown in panel (b) at $d_{\mathrm{NiFe}} \approx 2.1$ nm. That is the main result of this work.

For completeness, we mention that $I_c R_N$ data on NiFe junctions has been previously published by Robinson \textit{et al.} \cite{Robinson2006}, Dayton \textit{et al.} \cite{Dayton2018} and by Glick \textit{et al.} \cite{Glick2017}. The junctions in the latter study were fabricated and measured in our laboratory and exhibit $I_c R_N$ products similar to those shown in the right panel of Fig. \ref{fig:ICRN}, but with more scatter in the data.

To obtain values of $J_c$ from the data shown in Fig. \ref{fig:ICRN}, one should divide $I_c R_N$ by the area-resistance product, $AR_N$. In junctions with very thick F layers, one expects $AR_N$ to increase linearly with $d_F$ due to the resistivity of the F material \cite{Khaire2009}. Because the F layers are so thin in the junctions studied here, $AR_N$ is dominated by the interface resistance between the superconducting Nb electrodes and the F layer \cite{Fierz1990}. (The Cu spacer layers make no discernible difference.) Typical S/F boundary resistances lie in the range $AR_N$ = 5--\SI{8}{\femto \ohm \meter^2}, which for a junction with area of \SI{0.5}{\micro \meter^2} translates to a resistance of 10--\SI{16}{\milli \ohm}. Our junctions have resistances that cluster around 16--\SI{21}{\milli \ohm} for most of the junctions -- see Fig. \ref{fig:WR} in Appendix A. We attribute these slightly higher resistance values to the extra interfaces associated with the 5-nm Nb layer deposited in the initial sputtering run, which serves to protect the F layers from oxidation during the subsequent processing steps. The important point is that the resistances of the junctions with the extra Ni layers, 19.3 $\pm$ 2.4 m$\Omega$, are similar to those for the reference junctions without Ni layers, 20.1 $\pm$ 4.9 m$\Omega$. So the factor of four enhancement of $I_c R_N$ shown in Figure 5 translates to an equivalent enhancement of $J_c$.

\section{Discussion}

The behavior of $I_c R_N$ versus ferromagnet thickness has been calculated theoretically \cite{Buzdin2005} and measured experimentally by many groups for a number of ferromagnetic materials \cite{Ryazanov2001, Kontos2002, Blum2002, Sellier2003}. The $I_c R_N$ versus F layer thickness function is predicted to oscillate and decay either algebraically for ballistic transport \cite{Buzdin1982} or exponentially for diffusive transport \cite{Buzdin1991}. For very weak ferromagnets with nearly identical majority and minority spin bands, the Usadel equations govern this oscillatory-decay behavior \cite{Buzdin2005}. However, those equations are not appropriate for strong ferromagnetic materials because they do not take into account the complex band structures of the materials and the mismatches of Fermi surfaces that occur at each interface. At the very least, the theoretical models should incorporate the differences in the densities of states and diffusion constants for majority and minority spins. More realistic microscopic calculations based on Density Functional Theory and the Bogulibov-deGennes equations have been performed for the special case of Nb/Ni/Nb Josephson junctions by Ness \textit{et al.} \cite{Ness2022}, but unfortunately such calculations have not yet been carried out for any other ferromagnetic material, or for junctions that contain more than one material.

\begin{table*}[!htbp]
    \centering
    \begin{tabular}{|c|c|c|c|c|}
\hline
Sample & $V_0$ (\SI{}{\micro \volt}) & $\xi_{F1}$ (nm) & $\xi_{F2}$ (nm) & $d_{0-\pi}$ (nm)\\
 \hline
Ni(0.4)/NiFe($d_F$-0.8)/Ni(0.4) &  $800\pm110$ & $0.64\pm0.02$ & $0.78\pm0.01$ & $1.00\pm0.02$\\
Ni(0.2)/NiFe($d_F$-0.4)/Ni(0.2) &  $349\pm39$ &  $0.76\pm0.05$ & $0.78$ (fixed) & $0.92\pm0.01$\\
NiFe($d_{\mathrm{NiFe}}$) & $252\pm48$ & $0.71\pm0.04$ & $0.74\pm0.06$ & $1.49\pm0.01$\\
 \hline
    \end{tabular}
    \caption{Parameters determined from fits of Eqn. \ref{Eqn:Diffusive} to the data shown in Fig. \ref{fig:ICRN} for junctions containing Ni(0.4)/NiFe($d_F$-0.8)/Ni(0.4), Ni(0.2)/NiFe($d_F$-0.4)/Ni(0.2) and NiFe($d_{\mathrm{NiFe}}$). We fabricated and measured fewer of the junctions with 0.2 nm Ni layers, so there are not enough data points to establish the oscillation period. Hence we fixed the value of $\xi_{F2}$ to the value obtained from the first data set, $0.78$ nm.}
    \label{tab:parameters}
\end{table*}

Due to the rather short mean free path of minority electrons in NiFe \cite{Vila2000} as well as the multiple NiFe/Cu or Ni/Cu interfaces in our junctions, we expect the transport through our junctions to be diffusive. Hence we fit our $I_c R_N$ vs thickness data to the following equation:
\begin{equation} \label{Eqn:Diffusive}
    I_c R_N = V_0\; \mathrm{exp}\left(\frac{-d_F}{\xi_{F1}}\right) \left| \mathrm{sin} \left( \frac{d_F - d_{0-\pi}}{\xi_{F2}} \right)  \right|
\end{equation}
where $V_0$ sets the overall magnitude of $I_c R_N$, $\xi_{F1}$ and $\xi_{F2}$ are length scales that control the decay and oscillation period in the ferromagnet F, and $d_{0-\pi}$ is the thickness where the first $0-\pi$ transition occurs. Even though Ni and NiFe have different values of $\xi_{F1}$ and $\xi_{F2}$, for simplicity we consider the Ni/NiFe/Ni trilayer as a single F layer instead of three separate ones. The solid lines in Fig. \ref{fig:ICRN} are fits of Eqn. \ref{Eqn:Diffusive} to the data with experimental uncertainties determined from the Airy function fits discussed previously. (The uncertainties are smaller than the symbol size in Fig. \ref{fig:ICRN}.) The fit parameters for all three data sets are tabulated in Table \ref{tab:parameters}.

The enhancement of the critical current in the $\pi$-state of NiFe Josephson junctions has been achieved by adding 0.4 nm of Ni at each NiFe/Cu interface. But how does this ``interface engineering" actually work? We offer several perspectives on this question. Conceptually, one may think of this effect as analogous to impedance matching of transmission lines. The major difference, of course, is that the impedance is a single number, whereas at a solid-solid interface one has to sum the transmissions of the k-states over the entire Fermi surfaces of both materials at the interface. Using simplified models of interfaces that neglect details of band structure, Pugach \textit{et al.} \cite{Pugach2011} and Heim \textit{et al.} \cite{Heim2015} have shown theoretically that the properties of the S/F interface can have a strong influence on both the F-layer thickness where the 0-$\pi$ transition occurs and on the magnitude of the maximum critical current in the $\pi$ state. Indeed, both effects are present in the data shown in Fig. \ref{fig:ICRN} and in the fit parameters in Table \ref{tab:parameters}. Those indicate that the position of the first 0-$\pi$ transition has been shifted from an F-layer thickness of 1.5 nm for the case of NiFe to a total F-layer thickness of $0.9 - 1.0$ nm for the NiFe surrounded by thin Ni layers. As a result of that shift, the F-layer thickness corresponding to the maximum value of $I_c$ in the $\pi$-state is reduced from about 2.1 nm to about 1.6 nm in the junctions with the thin Ni layers. And since part of that 1.6 nm is pure Ni, there is substantially less NiFe in the junctions with the Ni layers compared to those without. So a simple explanation for the enhancement of $I_c$ is that there is less NiFe, hence less decay of $I_c$ due to NiFe.

Our own perspective on this issue is somewhat different, and derives from studies of Giant Magnetoresistance (GMR) of magnetic multilayers over the past 30 years \cite{Bass2016}. In particular, when measurements are performed in the current-perpendicular-to-plane (CPP) geometry, the magnetoresistance can often be calculated using a simple ``two-current series resistor'' (2CSR) model, which is valid when the spin diffusion length is much longer than the total thickness of the multilayer \cite{Zhang1991,Lee1993,Valet1993}. In the 2CSR model, the area-resistance product (AR) of the multilayer with arbitrary but collinear magnetization directions of the various ferromagnetic layers can be expressed in terms of the various layer thicknesses and a set of parameters characterizing the spin-dependent resistivities in the bulk materials and spin-dependent boundary resistances at the interfaces. For example, transport through an F/N interface is decribed by two interface specific resistances:  $AR^{\uparrow}_{F/N}$ and $AR^{\downarrow}_{F/N}$, where $\uparrow$ and $\downarrow$ indicate conduction electron moment pointing parallel or antiparallel to the F-layer magnetization, respectively. Alternatively one can use the following parameters: the dimensionless interface scattering asymmetry $\gamma_{F/N} = (AR^{\uparrow}_{F/N}-AR^{\downarrow}_{F/N})/(AR^{\uparrow}_{F/N}+AR^{\downarrow}_{F/N})$ and twice the enhanced interface specific resistance $2AR^*_{F/N}= (AR^{\uparrow}_{F/N} + AR^{\downarrow}_{F/N})/2$. Clearly a large interface resistance is indicative of a low average transmission probability for electrons crossing an interface. Since the Cooper pairs coming from the conventional superconducting electrodes consist of electrons with opposite spins, a strong interfacial scattering asymmetry indicates poor transmission of one spin species across the interface. Hence we suggest that supercurrent will tend to be large in systems with small values of both $2AR^*$ and $\gamma$. (To achieve large GMR, on the other hand, one generally wants large values of those quantities.) The parameters for the interfaces discussed in our paper are tabulated in Table \ref{tab:interface}. Indeed, the Cu/Ni interface has lower values of both quantities than the Cu/NiFe interface. We acknowledge that the
advantage of lower $2AR^*$ and $\gamma$ might be partially offset by the addition of the two Ni/NiFe interfaces to the junctions; given how thin the Ni layers are, however, it is plausible that one can view the Ni/NiFe/Ni trilayer as a single material with a spatial gradient in the Fe concentration, rather than as a trilayer of three distinct materials. 

\begin{table}[!htbp]
    \centering
    \begin{tabular}{|c|c|c|}
\hline
 F/N Interface & $\gamma_{F/N}$ & $2AR^*_{F/N}$ (\SI{}{\femto \ohm \meter^2})\\
 \hline
 NiFe/Cu &  0.7 & 1.0\\
 Ni/Cu & 0.3 & 0.36 \\
\hline
    \end{tabular}
    \caption{Interface spin scattering asymmetry $\gamma_{F/N}$ and interface resistance $2AR^*_{F/N}$ for NiFe/Cu and Ni/Cu interfaces obtained from GMR studies \cite{Bass2016}.}
    \label{tab:interface}
\end{table}

As a final note, we mentioned in the Introduction that one can achieve high $I_cR_N$ product in ferromagnetic junctions with structure S/I/s/F/S. For example, the junctions reported in Refs. \cite{Ryazanov2012, Larkin2012, Vernik2013, Caruso2018} achieve values of $I_cR_N$ of about 0.7 mV -- not much smaller than that in Nb-based S/I/S junctions. The critical current densities of those junctions, however, are quite small: $J_c \approx 4.5$ kA/cm$^2$, so they would not be appropriate for applications requiring large $J_c$. In contrast, the critical current densities of our junctions in the $\pi$ state are about 400 kA/cm$^2$.

\section{Conclusion}
In conclusion, supercurrent transmission through NiFe Josephson junctions can be enhanced by adding thin layers of Ni at the interface on both sides. The cost to magnetic switching properties is minimal for samples with Ni(0.2) and Ni(0.4) layers. For samples with Ni(0.4) layers, there was roughly a factor of 4-5 improvement in the supercurrent transmission in the $\pi$-state. We suggest that this is due to better band-matching at the Cu/Ni interface versus the Cu/NiFe interface. First-principles band-structure calculations coupled with calculations of Josephson junction supercurrent, similar to those performed in ref. \cite{Ness2022}, would be helpful to study this effect quantitatively and establish the exact mechanism.


\begin{acknowledgments}
The authors thank V. Aguilar, T.F. Ambrose, M.G. Loving, A.E Madden, D.L. Miller, and N.D. Rizzo for helpful discussions. N.O. Birge acknowledges many years of stimulating discussions with J. Bass and W.P. Pratt, Jr. We also thank B. Bi and D. Edmunds for technical assistance, and we acknowledge use of the W. M. Keck Microfabrication Facility. This research was supported by Northrop Grumman Corporation.
\end{acknowledgments}

\appendix
\section{London penetration depth and process variation}
\begin{figure}[!htbp]
\includegraphics[width=\linewidth]{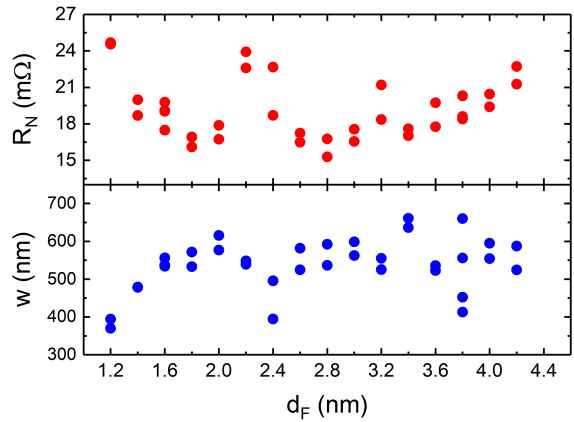}
\caption{Normal state resistance $R_N$ (red) and junction width $w$ (blue) vs total F-layer thickness $d_F$ for Ni(0.4)/NiFe($d_F$-0.8)/Ni(0.4).}
\label{fig:WR}
\centering
\end{figure}

Theoretical calculations show that the effective London penetration depth $2\lambda_{\mathrm{eff}}$ in Eqn. \ref{Eqn:Flux} is given by:
\begin{equation} \label{Eqn:London}
   2\lambda_{\mathrm{eff}} =  \lambda_1 \mathrm{tanh} \left(\frac{d_1}{2\lambda_1}\right) + \lambda_2 \mathrm{tanh} \left(\frac{d_2}{2\lambda_2}\right)
\end{equation}
where $\lambda_1$, $\lambda_2$ are the London penetration depths and $d_1$, $d_2$ are the thicknesses of the top and bottom leads, respectively \cite{BaronePaterno1982}.  That expression approaches $\lambda_1+\lambda_2$ when both electrodes are much thicker than their respective London penetration depths, or $(d_1+d_2)/2$ in the opposite limit of thin electrodes. Our samples are in the intermediate case, with each electrode thickness within a factor of two of its $\lambda$ value. We discovered recently \cite{Madden2018,Quarterman2020} that $\lambda$ for our bottom [Nb/Al]$_3$/Nb multilayer is in the vicinity of 185 nm -- much longer than the value of 85 nm we expect for our sputtered Nb films \cite{Khaire2009}. If we use $\lambda$ = 185 nm and 85 nm for the bottom and top electrodes, respectively, and evaluate Eqn. \ref{Eqn:London} for our samples with $d_1 = 102.2$ nm and $d_2 = 150$ nm, we obtain the result $2\lambda_{\mathrm{eff}} \approx 110$ nm. To calculate the magnetic flux through through the junction due to the external field, we must add the total thickness of the normal layers, $d_N = 14$ nm, and the thickness of the ferromagnetic layers, $d_F$. Using the above parameters, we can extract values of the junction widths, $w$, from the fits of Eqn. \ref{Eqn:Airy} to the experimental Airy patterns. Those values are plotted in the lower panel of Fig. \ref{fig:WR} for the complete set of Ni(0.4)/NiFe($d_F$-0.8)/Ni(0.4) junctions. The average junction width is about 550 nm, which agrees with estimates obtained from scanning electron microscopy pictures taken during the fabrication process.

The data in Fig. \ref{fig:WR} show considerable scatter in the junction widths derived from the Airy patterns. In the upper panel of the figure, we plot the normal state resistance $R_N$ of the junctions. There is a noticeable inverse correlation between these two, especially in the extreme cases. Ideally, we shouldn't expect such a high degree of scatter in the junction widths, since the e-beam lithography process is highly reproducible and the ion milling rate typically varies by only a few percent during the 2-minute milling procedure. We suspect that the culprit is the thermal deposition of $\mathrm{SiO_x}$, which is the least well-controlled process. The color of the deposited $\mathrm{SiO_x}$ layer varies between different shades of brown which indicates a variability of thickness. This variation in $\mathrm{SiO_x}$ thickness can lead to variations in the lateral size of the junctions, which leads to varying resistances. Optical microscopy inspection of our junctions during the fabrication process reveals that the junctions with higher $R_N$ values tend to have smaller junction openings after lift-off of the negative e-beam resist. This is consistent with those junctions having smaller extracted junction widths $w$. 

For the data shown in Fig. \ref{fig:WR}, the mean and standard deviation of the resistance measurements is 19.2 $\pm$ 2.5 m$\Omega$, whereas for the widths we find 551.2 $\pm$ 69.6 nm.  In relative terms, the resistances vary by $\pm$ 13\% while the widths vary by $\pm$ 12.6\%. Since resistance is inversely proportional to junction area, one would expect larger relative variations in resistance; the fact that that is not the case indicates that other unknown factors contribute to the apparent width variation deduced from the period of the Airy pattern.  

\section{Airy pattern field shifts}

\begin{figure}[!htbp]
\includegraphics[width=\linewidth]{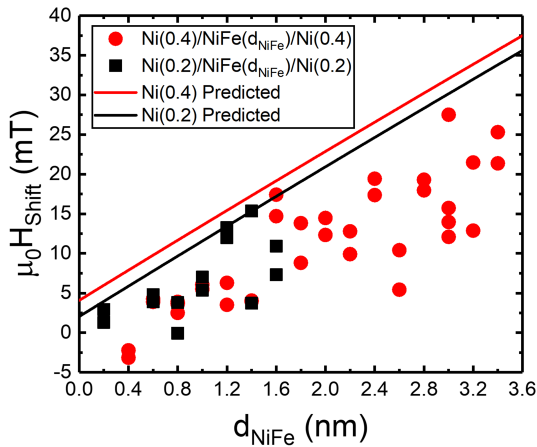}
\caption{Field shift vs NiFe thickness for Ni(0.4)/NiFe($d_{\mathrm{NiFe}}$)/Ni(0.4) (red circles) and Ni(0.2)/NiFe($d_{\mathrm{NiFe}}$)/Ni(0.2) (black squares). Each data point represents the average field shift of the upsweep and downsweep data, taking into account their opposite signs. Solid lines are predictions based on Eqn. \ref{Eqn:Hshift} as discussed in the text. The fact that the data points fall below the lines may indicate that the remanent magnetic states inside the junctions are not fully saturated.}
\label{fig:Hshift}
\centering
\end{figure}

As discussed earlier, the data shown in Fig. \ref{fig:Fraunhofers} are shifted in field due to the contributions of the Ni and NiFe magnetizations to the total magnetic flux in the junction. If the magnetic trilayer is uniformly magnetized at remanence, and if we can neglect any magnetic flux from the F layer that returns inside the junction, then Eqn. \ref{Eqn:Flux} predicts the field shift to be
\begin{equation} \label{Eqn:Hshift}
    \mu_0 H_{\mathrm{shift}} =  \frac{-\mu_0 (M_{\mathrm{Ni}} d_{\mathrm{Ni}}+ M_{\mathrm{NiFe}} d_{\mathrm{NiFe}})}{(2\lambda_{\mathrm{eff}} + d_N + d_{\mathrm{Ni}}+d_{\mathrm{NiFe}})}
\end{equation}
where we have replaced $M d_F$ in Eqn. \ref{Eqn:Flux} with $M_{\mathrm{Ni}}d_{\mathrm{Ni}} + M_{\mathrm{NiFe}}d_{\mathrm{NiFe}}$. The field shift values obtained by fitting the Fig. \ref{fig:Fraunhofers} data to Eqn. \ref{Eqn:Airy} are plotted vs NiFe thickness in Fig. \ref{fig:Hshift} for the Ni(0.4)/NiFe($d_{\mathrm{NiFe}}$)/Ni(0.4) and Ni(0.2)/NiFe($d_{\mathrm{NiFe}}$)/Ni(0.2) sample sets, as the red and black symbols, respectively. There is significant scatter in the data; nevertheless, the field shift does increase with increasing NiFe thickness, as expected. The solid lines are the predicted values calculated from Eqn. \ref{Eqn:Hshift} using the values of $M_{\mathrm{Ni}}$ and $M_{\mathrm{NiFe}}$ from the magnetic data in Fig. \ref{fig:Msat_NiNiFe}, and the value of $2\lambda_{\mathrm{eff}} + d_N = 124$ nm as discussed in Appendix A. The data points all fall below the lines, indicating either that there is a substantial amount of flux from the F layer that returns inside the junction, or that the remanent magnetization of the Ni/NiFe/Ni trilayers is less than the saturation magnetization. The latter might be due either to some magnetic relaxation at the ends of the ellipses, or to domain formation. Tolpygo {\it et al.} have performed similar measurements for junctions containing Ni and found the remanent magnetization to be only around 1/7th of the saturation magnetization \cite{Tolpygo2019}. That very low value is probably due to two factors: the measured junctions were rather large (circular diameter = \SI{2.24}{\micro \meter}) and the magnetizing field was quite low (10 mT); hence the Ni inside the junction was undoubtedly in a multidomain state with low net magnetization.

\bibliography{references}

\end{document}